\documentclass[11pt]{article}

\usepackage[final]{acl}

\usepackage{times}
\usepackage{latexsym}

\usepackage[T1]{fontenc}

\usepackage[utf8]{inputenc}

\usepackage{microtype}

\usepackage{inconsolata}

\usepackage{graphicx}
\usepackage{booktabs}
\usepackage{amsmath}
\usepackage{subcaption}
\usepackage{multirow}
\usepackage{xurl}

\title{Pruning for Efficiency, Paying in Fairness: Demographic Disparities in Pruned Speech-LLMs}

\author{
  \textbf{Ganesh Pavan Kartikeya Bharadwaj Kolluri},
  \textbf{Michael Kampouridis},
  \textbf{Ravi Shekhar}
\\
  School of Computer Science and Electronic Engineering, University of Essex, UK
\\
  \texttt{\{karthik.kolluri, mkampo, r.shekhar\}@essex.ac.uk}
}

\begin{document}
\maketitle
\begin{abstract}

Speech-LLMs are expensive to run, making compression important for real-world deployment. However, compressed models are usually selected using aggregate word error rate (WER), which can hide how pruning affects different demographic groups. In this work, we systematically study the effect of audio encoder pruning on SLAM-ASR for different demographic groups. Using the Fair-Speech and Common Voice datasets, we found that the pruning does not affect all demographic groups equally; the gap between best- and worst-performing groups increases in fold.  These disparities appear across all three encoder scales, but only the largest model initially hides them behind aggregate WER. LoRA adaptation improves WER for every group, but benefits groups already performing well more strongly and widens for certain groups. On Common Voice English, Danish, and Dutch, accent gaps persist but do not clearly widen, showing that the fairness effects of pruning vary across datasets and must be measured directly. Our findings suggest that for pruned models, deployment decisions should include per-group WER, with the worst-performing group's error rate as an explicit criterion.

\end{abstract}

\section{Introduction}

Automatic speech recognition (ASR) transcribes spoken language, and it is used in voice assistants, dictation tools, and captioning services~\citep{Koenecke2020RacialDI}. The field's current dominant architecture is the Speech-LLM: a pretrained speech encoder coupled to a large language model through a lightweight projector. These stacks are generally large, making them computationally expensive, and model compression has become a routine step between training and deploying. 

\begin{figure}[ht]
\centering
\includegraphics[width=\columnwidth]{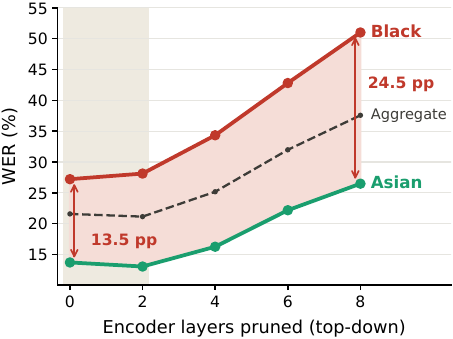}
\caption{\textbf{Compression widens the racial gap, and the aggregate WER hides it.} WER for Black and Asian speakers (Fair-Speech, Whisper large) under top-down pruning; the wedge between the curves is their gap, which nearly doubles ($13.5\!\rightarrow\!24.5$\,pp). Within the grey band, aggregate WER \emph{improves} ($21.6\!\rightarrow\!21.1$\%), yet Black speakers are the only group whose word error rate rises significantly ($+0.9$\,pp, $p{=}.009$).}
\label{fig:hook}
\end{figure}

One of the widely used compression techniques is pruning. Usually, a pruned model is validated on aggregate word error rate (WER). However, ASR systems are already performing unevenly across demographic groups~\citep{Koenecke2020RacialDI}. That means an aggregated WER hides group disparities. In that case, when we apply pruning to those models, the question will be: what does pruning do to that unevenness? \footnote{Code is available at \url{https://github.com/KarthikKolluriKB/SpeechLLM-Pruning-Fairness}}

We study this effect systematically in the SLAM-ASR pipeline \citep{ma2024embarrassingly, zhang-etal-2026-speak, zhang-etal-2026-language}: a frozen Whisper encoder~\cite{Radford2022RobustSR}, a trainable projector, and a frozen Qwen2.5-3B decoder~\cite{qwen2025qwen25technicalreport} connected together as an ASR pipeline. We experiment on three Whisper scales (12, 24, and 32 encoder layers), prune each encoder top-down two layers at a time, and retrain the projector at every depth, so each configuration is a deployable model. This pipeline is demographically blind: no group label enters training, pruning, or LoRA adaptation. Our primary corpus is Fair-Speech \citep{Veliche2024TowardsMF}, which offers the richest demographic schema, and we add Common Voice English, Dutch, and Danish for cross-lingual study. 

As shown in Fig.~\ref{fig:hook},
in a Speech-LLM built on Whisper Large, removing the top two encoder layers improves aggregate WER from 21.6 to 21.1 ($p{=}.038$). The same setting raises WER for Black speakers by 0.9 points ($p{=}.009$). Here, the aggregate WER hides the subgroup degradation: the total improves while a group gets worse. Motivated by this, in this work we address three research  questions:

\begin{itemize}
\item \textbf{RQ1:} Does encoder pruning amplify demographic
disparities, and does aggregate WER hide it?
\item \textbf{RQ2:} Does the effect depend on encoder scale and
language resource level?
\item \textbf{RQ3:} Does LoRA adaptation recover performance equally across groups?
\end{itemize}

Our results answer these questions as follows. Pruning amplifies the racial disparity on the largest model (RQ1). The gap between Black and Asian speakers grows from 13.5 to 24.5 percentage points (pp)  by eight removed layers, even though the first prune improves aggregate WER. The disparity exists at every encoder scale: on all three models, the worst-performing group starts with roughly twice the WER of the best-performing. Concealment, however, appears only at the largest scale (RQ2). On the small and medium models, the first prune raises aggregate WER by 10.8 and 2.3 pp, respectively. So the damage is already visible in the aggregate. Across languages, the accent gaps on Common Voice English and Dutch persist under pruning but do not amplify. Finally, LoRA improves aggregate WER at every depth but widens the Black-to-Asian ratio at every matched depth (RQ3): it recovers accuracy for every group, but better for the groups the model already performs well on and worse for the groups it performs worse on.

Our findings carry a direct practical message: aggregate WER cannot validate a pruned speech model as fair. Deployment decisions for compressed speech models should therefore include per-group WER, with explicit attention to the worst-performing group. To our knowledge, this is the first systematic study of how structural encoder pruning in Speech-LLM affects demographic disparity across pruning depth, model scale, and LoRA adaptation.

\section{Related work}

\subsection{Bias in ASR}

Performance disparities across speaker groups are well documented in ASR. \citet{Koenecke2020RacialDI} found that five commercial systems transcribed speech from Black speakers with an average word error rate (WER) of 35\%, against 19\% for white speakers, and traced the gap to the underlying acoustic models, since it persisted on a matched subset of identical phrases. Subsequent work investigated the linguistic mechanisms behind this disparity, linking errors to morpho-syntactic features of African American English such as habitual ``be'' \citep{Martin2020UnderstandingRD} and to ethnicity-related dialectal variation more broadly \citep{Wassink2022UnevenSA}. Disparities along gender and dialect lines were reported earlier \citep{Tatman2017GenderAD, Tatman2017EffectsOT}, and \citet{Feng2021QuantifyingBI} systematically quantified bias across gender, age, regional accent, and non-native speech; we adopt this per-group WER evaluation. These imbalances persist in large pretrained models: evaluations of Whisper \citep{Radford2022RobustSR} report higher accuracy for North American than for British or Australian accents, and for native than for non-native speech, with WER associated with speaker sex, first-language typology, and second-language proficiency \citep{Graham2024EvaluatingOW}. Performance gaps for child speakers likewise persist across Whisper scales \citep{Attia2023KidWhisperTB}. To measure such gaps systematically, corpora with speaker demographics have emerged: Common Voice collects optional self-reported accent, gender, and age at scale \citep{Ardila2019CommonVA}, the Edinburgh corpus targets accent diversity \citep{Sanabria2023TheEI}, and Fair-Speech provides the richest schema of the three, covering age, gender, ethnicity, socio-economic status, and geographic variation \citep{Veliche2024TowardsMF}. We justify our choice among these in Section~\ref{sec:setup}.

\subsection{Model Compression, Scale, and Bias}

How model compression affects bias is still an open question, with outcomes that vary by method, domain, and demographic axis. In computer vision, \citet{Hooker2020CharacterisingBI} demonstrated that pruning and quantization can amplify bias: minimal changes in overall accuracy may hide substantial errors affecting a small subset of under-represented examples, an observation that supports evaluating worst-group rather than average performance \citep{Sagawa2019DistributionallyRN}. \citet{Tran2022PruningHA} attributed such disparities to differences in gradient norms and decision-boundary distance across groups, while \citet{Iofinova2023BiasIP} found that, although moderate pruning need not increase bias, disparities for specific protected attributes emerge at high sparsity. %

In LLMs, compression's effect on fairness depends on the base model and method \citep{Ramesh2023Comparative}, but two direct comparisons find pruning riskier than quantization, for protected groups \citep{Xu2024BeyondPerplexity} and for trustworthiness \citep{Hong2024DecodingTrust}, and compression's effect on social bias varies with model scale \citep{Gonalves2023UnderstandingTE}.

In speech, evidence on compression and bias is scarce. \citet{Lin2024OnTS} examined self-supervised models under pruning and distillation, finding, in embedding-level probes, that distillation increased gender bias while decreasing age bias, whereas row pruning reduced bias. For speech-LLMs, bias studies either examine response content \citep{Lin2024ListenAS, Lin2024SpokenSO} or benchmark WER disparities across architectures: \citet{Ginjala2026DoLD} evaluated nine models spanning CTC, encoder-decoder, and LLM-decoder designs and reported that the audio encoder, more than language model scale, governs fairness.

We build directly on \citet{Kolluri2026OnTR}, who pruned Whisper encoder layers in the speech-LLM \citep{ma2024embarrassingly} and showed that removing a few layers costs little aggregate WER and that LoRA \citep{Hu2021LoRALA} recovers the unpruned baseline's WER; that study reports aggregate WER only. How adaptation affects fairness is itself unexplored: \citet{Ding2024OnFO} found no consistent subgroup effect of LoRA relative to full fine-tuning; low-rank adaptation can propagate non-demographic biases inherited from noisy pretraining data \citep{Chang2024BALoRABL}; and fine-tuning aimed at under-represented speakers lowers their WER \citep{Shekoufandeh2025ImprovingTI}, suggesting repair depends on where adaptation is aimed. No prior work examines whether encoder pruning shifts demographic disparity in a speech-LLM's transcriptions, or whether LoRA's recovery is shared evenly across groups; we address both.

\section{Experimental Setup}
\label{sec:setup}
\subsection{System and Pruning}

We follow the SLAM-ASR architecture~\cite{ma2024embarrassingly}: a Whisper encoder~\cite{Radford2022RobustSR} feeds a small trainable projector (ConcatLinear), which maps audio representations into a frozen Qwen2.5-3B decoder~\cite{qwen2025qwen25technicalreport}. We hold the LLM fixed so that encoder capacity is the only varied component; both the encoder and the LLM stay frozen, and only the projector and LoRA adapters are trained. To treat encoder capacity as an experimental variable, we run the full procedure with three Whisper variants, Small (12 layers), Medium (24), and Large-v2 (32), each paired with the same projector and LLM. As in~\citet{Kolluri2026OnTR}, we prune each variant top-down, removing two layers at a time, and report pruning as the absolute number of layers removed (L-2, L-4, \ldots). The projector is retrained at every depth, so each configuration is a fully retrained, deployable system rather than a fixed model probed after pruning.

\textbf{Procedure.} Concretely, the full protocol at each encoder scale (small, medium, and large-v2) proceeds as follows.
\begin{enumerate}
\itemsep0em
\item Initialise the system with the unpruned encoder and train the projector,
      recording the result as the $L$-0 baseline.
\item Remove the top two encoder layers to obtain depth $L$-$k$, for
      $k \in \{2, 4, 6, 8\}$.
\item Retrain the projector from scratch at that depth, so that each configuration
      constitutes a deployable system rather than a mismatched stack probed after
      pruning.
\item Select the configuration on aggregate WER alone; no demographic label enters
      training, pruning, or selection.
\item Decode the evaluation corpora and compute WER separately for each demographic
      group meeting the analysability threshold of Section~\ref{sec:eval-details}.
\end{enumerate}

The models never see demographic labels: training and model selection use aggregate WER alone. In practice, this is also how pruned models are judged, since compression work reports aggregate WER and stops there. Our question is what this number hides: does pruning damage the WER of individual demographic groups even when the aggregate looks fine? To answer it, we take each pruned model after training and evaluate it separately for every demographic group on the evaluation corpora.

\subsection{Datasets}

Table~\ref{tab:datasets} summarizes the corpora used for training and evaluation.

\textbf{Training data:} The projector and LoRA adapters are trained on Common Voice~22~\cite{Ardila2019CommonVA}, a crowdsourced multilingual read-speech corpus, following the same training configuration as \citet{Kolluri2026OnTR}. We use the official train and test splits without modification, so speakers are disjoint across splits by construction, and the counts reported in Table~\ref{tab:datasets} are those remaining after the duration filter described below. Our primary system uses the English subset (100 hours); for the cross-lingual analysis, we additionally train separate systems on the Danish (4.2 hours) and Dutch (54 hours) subsets, which span an order-of-magnitude range in training data.

\textbf{Evaluation data and demographic axes:} We aim to evaluate as many demographic axes as possible. To this end, we selected two datasets: Fair-Speech~\cite{Veliche2024TowardsMF} and  Common Voice~22~\cite{Ardila2019CommonVA}. Fair-Speech~\cite{Veliche2024TowardsMF} is a comprehensive evaluation benchmark built specifically to expose demographic disparities in speech recognition, and it carries the richest bias annotation of our corpora. Across 26{,}417 utterances, it labels \emph{ethnicity} (seven analyzable groups), \emph{socio-economic status}, gender, age, and first language. Fair-Speech is used exclusively for evaluation and contributes no training data. Common Voice~22~\cite{Ardila2019CommonVA} (CV-22) is a large multilingual crowdsourced corpus of read speech in which contributors may optionally self-report speaker attributes. We use this metadata to measure disparities by \emph{accent}, \emph{gender}, and \emph{age} on the English, Danish, and Dutch splits, which allowed us to compare results across languages. Because these attributes are optional, group-level results on CV-22 are computed over the subset of test utterances carrying the relevant annotation, which is smaller than the full test split. The dataset statistics and supported demographic axes are presented in Table~\ref{tab:datasets} and Table~\ref{tab:eval_data} respectively.

\begin{table}[t]
\centering

\setlength{\tabcolsep}{6pt}
\renewcommand{\arraystretch}{1.2}
\small
\begin{tabular}{llr}
\toprule
\textbf{Corpus} & \textbf{Split} & \textbf{Samples} \\
\midrule
\multicolumn{3}{l}{\textit{Training (Common Voice 22)}} \\
\midrule
English (EN) & Train & 58{,}140 \\
Dutch (NL)  & Train & 43{,}458 \\
Danish (DA) & Train & 3{,}592  \\
\midrule
\multicolumn{3}{l}{\textit{Evaluation}} \\
\midrule
Common Voice (EN) & Test & 16{,}391 \\
Common Voice (NL) & Test & 12{,}033 \\
Common Voice (DA) & Test & 2{,}684  \\
Fair-Speech       & eval & 26{,}417 \\
\bottomrule
\end{tabular}
\caption{Dataset statistics. Common Voice 22 provides the training and test splits; Fair-Speech is used for evaluation only.}
\label{tab:datasets}
\end{table}

\begin{table}[t]
\centering

\setlength{\tabcolsep}{4.3pt}
\renewcommand{\arraystretch}{1.45}
\small
\begin{tabular}{@{}ll@{}}
\toprule
\textbf{Corpus} & \textbf{Bias axes} \\
\midrule
CV-22 (EN/DA/NL) & accent, gender, age \\
Fair-Speech & ethnicity, SES, gender, age, L1 \\
\bottomrule
\end{tabular}
\caption{Evaluation corpora and the demographic axes each supports.}
\label{tab:eval_data}
\end{table}

\textbf{Preprocessing.} Audio is resampled to 16\,kHz and converted to 80-channel
log-Mel spectrograms following the Whisper pipeline. We retain utterances between
0.5 and 30 seconds. Transcriptions are lowercased with punctuation removed,
preserving apostrophes for contractions.

\subsection{Implementation Details}
\label{sec:impl}

\textbf{Model Architecture.} Our system follows SLAM-ASR~\cite{ma2024embarrassingly} and comprises three components: a Whisper speech encoder~\cite{Radford2022RobustSR}, a lightweight ConcatLinear projector, and a Qwen2.5-3B LLM~\cite{qwen2025qwen25technicalreport}. We run it with three Whisper encoder variants, Small (12 layers), Medium (24), and Large-v2 (32), each paired with the same projector and LLM. 

\textbf{Projector.} The projector is a two-layer MLP with concatenation-based temporal downsampling: it concatenates five consecutive frames, reducing the sequence length five times, then applies two linear layers with ReLU and dropout 0.1 (hidden dimension 2048), followed by LayerNorm to match the scale of the LLM text embeddings. The projector is reinitialized and retrained from scratch at every pruning depth rather than fine-tuned from the unpruned checkpoint.

\textbf{LoRA.} We apply LoRA~\cite{Hu2021LoRALA} to the query, key, value, and output projections of all LLM attention layers. To limit overfitting on smaller datasets, we set the rank per resource level: $r{=}8$, $\alpha{=}16$ for Danish (low-resource) and $r{=}16$, $\alpha{=}32$ for Dutch and English; $r{=}16$ on Danish overfit in preliminary runs, motivating the reduced rank. LoRA modules use dropout 0.1, adding roughly 0.8M ($r{=}8$) and 1.5M ($r{=}16$) trainable parameters.

\textbf{Training Details.} All models are trained with AdamW~\cite{loshchilov2017decoupled} at learning rate $1\times10^{-4}$, weight decay 0.01, and gradient clipping 1.0, under a cosine schedule with linear warmup over the first 5\% of steps, in bfloat16 on a single NVIDIA A6000 (48\,GB). The Whisper encoder and Qwen2.5-3B weights remain frozen; only the projector and, where enabled, the LoRA adapters are updated. For English configurations, we use a batch size of 8 and train for 2 epochs, or 4 with LoRA. Dutch uses a batch size of 8 and Danish a batch size of 4, with more passes over the smaller corpora; per-configuration epoch counts are given in Appendix~\ref{app:repro}. Training stops early on validation WER, so model selection uses aggregate validation performance alone, with no demographic label involved. Every pruning depth within a setting uses the same setup, so depths differ only in encoder size. All runs use a single seed (42), so we rely on the paired bootstrap in Section~\ref{sec:eval-details} rather than variance across seeds.

\textbf{Decoding.} At inference, the projected audio embeddings are prepended to a fixed plain-text prompt, \texttt{Transcribe speech to text.}, applied without a chat template and held constant across all scales, depths, and languages. Transcriptions are generated with beam search (width 2, no sampling, no repetition or length penalty) up to 128 new tokens, identically for base and LoRA-adapted systems. Checkpoint selection during training uses greedy decoding on the validation split with a repetition penalty of 1.5.

\textbf{Baselines.} The unpruned ($L$-0) system at each encoder scale is the reference for every pruned configuration at that scale, so configurations differ only in encoder depth and in the projector retrained at that depth. LoRA comparisons are matched the same way, against the base system at the same scale and depth. We report no external ASR baseline, since our claim concerns how a fixed pipeline behaves under compression rather than how it compares against other systems.

\subsection{Evaluation Details}
\label{sec:eval-details}
We evaluate every configuration with WER: aggregate WER is computed over the full evaluation set, and a group's WER over the utterances spoken by members of that group, so every group is scored by the same rule.

We assess disparity with three measures, fixed before the analysis: the worst-group WER at each depth as our primary measure, the absolute gap $\Delta = \mathrm{WER}_{\text{worst}} - \mathrm{WER}_{\text{best}}$ in percentage points (pp), and the disparity ratio $\rho = \mathrm{WER}_{\text{worst}}/\mathrm{WER}_{\text{best}}$ as a scale-free check; within a group we report relative degradation $\mathrm{WER}_{Lx}/\mathrm{WER}_{L0}$. Since $\Delta$ and $\rho$ diverge by construction when all groups degrade together, we report both at every depth, tracking the Black--Asian pair throughout as the worst- and best-served groups. Significance comes from a paired resampling test: we draw 2{,}000 bootstrap resamples of the utterances, re-score both setups on each, and take $p$ as the fraction in which the WER difference flips direction. We limit our claims to the usable range, the contiguous depths at which aggregate WER stays at or below 40\%, and analyze groups only where they have at least 200 utterances and 30 minutes of audio; smaller groups appear in tables for coverage only.

\section{Results and Discussions}
\label{sec:results}
\subsection{Pruning and demographic disparity (RQ1)}
\label{sec:results-rq1}

\begin{table}[t]
\centering
\footnotesize
\setlength{\tabcolsep}{4pt}
\begin{tabular}{lrccccc}
\toprule
Group & $n$ & L-0 & L-2 & L-4 & L-6 & L-8 \\
\midrule
Asian       & 3,854 & $13.7$ & $13.1$ & $16.3$ & $22.2$ & $26.5$ \\
Native Haw. &   969 & $13.4$ & $13.7$ & $17.8$ & $24.7$ & $31.4$ \\
Hispanic    & 2,811 & $20.0$ & $19.2$ & $22.5$ & $27.5$ & $33.9$ \\
White       & 5,619 & $20.6$ & $20.1$ & $22.7$ & $28.4$ & $31.7$ \\
Native Am.  & 4,616 & $21.5$ & $19.0$ & $22.1$ & $29.0$ & $33.5$ \\
MENA        &   749 & $23.0$ & $22.3$ & $26.0$ & $34.2$ & $37.3$ \\
Black       & 7,799 & $27.2$ & $28.1$ & $34.3$ & $42.8$ & $51.0$ \\
\midrule
Aggregate   & 26,417 & $21.6$ & $21.1$ & $25.2$ & $32.0$ & $37.6$ \\
\bottomrule
\end{tabular}
\caption{Fair-Speech, ethnicity (Whisper large); $n$ is the number of utterances per group. Black speakers have the highest WER at every depth. At L-2 the aggregate improves while Black speakers degrade, and by L-8
the Black--Asian gap has nearly doubled, from 13.5 to 24.5\,pp.}
\label{tab:wer-eth}
\end{table}

We present our results by demographic dimensions, starting with ethnicity. Table~\ref{tab:wer-eth} presents WER for the seven ethnic groups as we prune Whisper large encoder layers: at L-0 (without pruning), the groups are already far apart, from 13.4\% for Native Hawaiian and 13.7\% for Asian speakers to 27.2\% for Black speakers, twice the Asian rate. The bias already exists before any pruning.

As we prune the top two encoder layers, the aggregate WER shows that the model performs slightly better: it falls from 21.6\% to 21.1\% ($p{=}.038$, Table~\ref{tab:wer-eth}). The group columns tell a different story: Black speakers are the only group whose error rate rises among other groups ($+0.9$\,pp, $p{=}.009$), Native American speakers improve most ($-2.5$\,pp, $p{=}.001$), and no other group changes significantly ($p \ge .13$). In relative terms, L-2 is closer to baseline on every group (aggregate at 0.98) except the Black group, at 1.03 (Appendix~\ref{app:rel}). A practitioner relying on aggregate WER alone would see a small win, while the group that started worst has become worse. 

At deeper prunes, all groups degrade but at different rates. At L-4 the aggregate has risen to 25.2\%, a cost any practitioner would notice, and the groups have begun to separate: Asian speakers are at 16.3\% while Black speakers are at 34.3\%, a gap of 18.0\,pp. At L-6 the aggregate is at 32.0\%, within the 40\% usability bound, while Black speakers are at 42.8\% and Asian speakers are at 22.2\%. The same configuration is therefore usable by the aggregate criterion and unusable for its worst-performing group. At L-8, Black speakers are at 51.0\% and Asian speakers at 26.5\%. Over the full sweep, Black speakers lose 23.8\,pp and Asian speakers 12.8\,pp, so the gap between them widens from 13.5\,pp to 24.5\,pp. Aggregate WER is misleading in two ways: it reports an improvement at the first prune (L-2) while the worst-performing group degrades, and it stays within the usability bound at depths where the worst-performing group has already exceeded it.

\begin{figure}[t]
\centering
\includegraphics[width=\columnwidth]{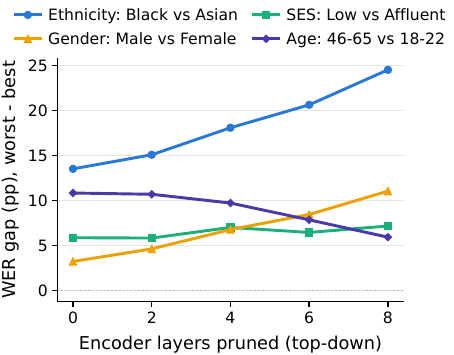}
\caption{Worst-best WER gap per demographic axis under top-down pruning (Fair-Speech, Whisper large); one line per axis, with the pair named in the legend. The ethnicity and gender gaps widen, the Socioeconomic gap holds, and the age gap narrows by leveling down.}
\label{fig:gaps}
\end{figure}

\begin{figure*}[!htb]
\centering
\includegraphics[width=\textwidth]{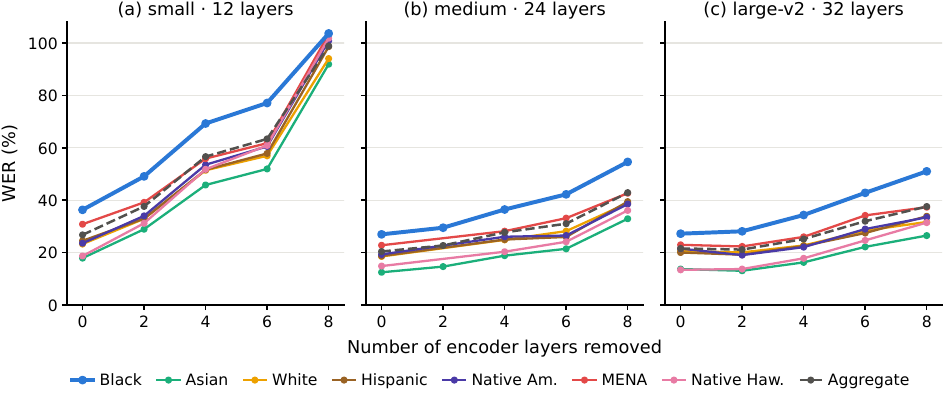}
\caption{Per-group WER on Fair-Speech as encoder layers are removed, per Whisper scale (shared $y$-axis; dashed = aggregate). Black speakers have the highest error rate at every scale and every depth. Only at large-v2 does the shallow prune leave the aggregate unchanged. Appendix Table~\ref{tab:rel-scale} covers that depth.}
\label{fig:scale}
\end{figure*}

Ethnicity is not the only axis that moves. Fig.~\ref{fig:gaps} plots the worst-to-best gap for each demographic axis at every depth (per-group values in Table~\ref{tab:app-axes}), and the three remaining axes each do something different. The gender gap grows from 3.3 to 11.1\,pp (male speakers disadvantaged in this corpus; Fig.~\ref{fig:app-gender}). The age gap narrows, but only because the oldest group, worst at baseline, degrades slowest (Fig.~\ref{fig:app-age}). The socioeconomic gap holds steady (Fig.~\ref{fig:ses}; Table~\ref{tab:app-axes}): Low-SES speakers start 5.9\,pp behind Affluent speakers (ratio 1.36) and stay behind at every depth, and although the raw gap reaches 7.2\,pp at L-8, Affluent speakers lose more of their starting accuracy (+83\% vs +67\%), so the ratio falls to 1.24. Pruning carries the class disparity along unchanged.

Taken together, these results answer RQ1. The disparity is inherited, but the hiding is new: at L-2 the aggregate improves while the worst-performing ethnicity group degrades significantly. Gender amplifies, class persists, age narrows by leveling down; aggregate WER is blind to all three, which is why pruned models need per-group evaluation.

\begin{figure*}[!htb]
\centering
\includegraphics[width=\textwidth]{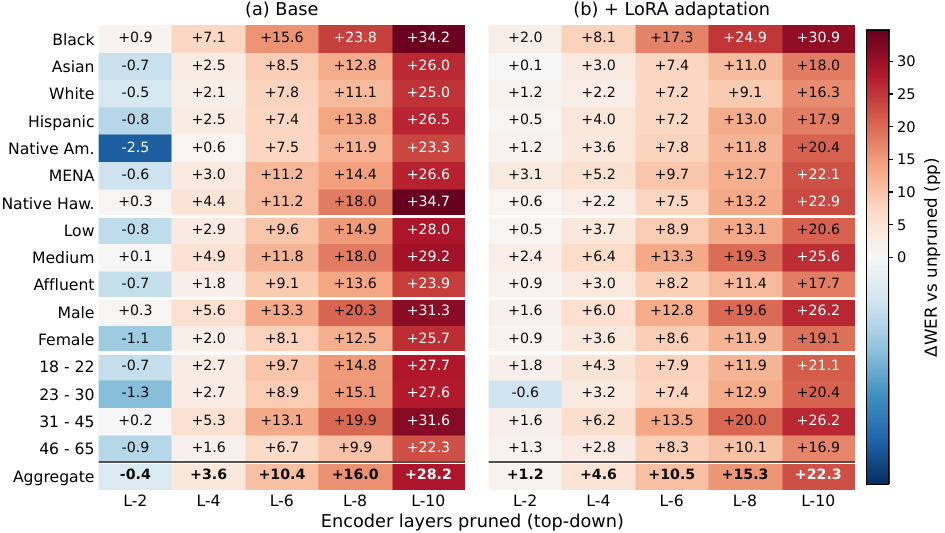}
\caption{Change in WER (percentage points) per subgroup as Whisper large-v2 is pruned top-down, without (a) and with (b) LoRA (Fair-Speech). Warm colors indicate degradation, cool colors indicate improvement, and the bottom row is the aggregate. Black speakers degrade most at nearly every depth in both conditions. LoRA lowers the aggregate everywhere, but it improves only the already better-performing groups: at ten removed layers it recovers $8.0$\,pp for Asian speakers and only $3.3$\,pp for Black speakers.}
\label{fig:heatmap-pair}
\end{figure*}

\subsection{The role of encoder scale (RQ2)}
\label{sec:results-scale}

Fig.~\ref{fig:scale} shows per-group WER at all three encoder scales. Black speakers have the highest WER in every panel at every depth. What differs across panels is the cost of the first prune. The small model degrades immediately, its aggregate rising by 10.8\,pp at two removed layers, and medium by 2.3\,pp; large-v2 instead improves by 0.5\,pp. 

The disparity ratio $\rho$ (Table~\ref{tab:rho-scale}) separates the three scales in a way the aggregate does not. Before pruning, they are similarly unequal, with $\rho$ at 1.99 on large-v2, 2.03 on small, and 2.16 on medium, so Black speakers face roughly twice the word error rate of Asian speakers on all the encoder scales. The first prune is what shows them going apart. On small and medium, the ratio falls, from 2.03 to 1.70 and from 2.16 to 2.01, because these models degrade so sharply that the better-served group loses more in proportional terms. On large-v2 it rises instead, from 1.99 to 2.15, and large-v2 is the one configuration whose aggregate WER improved.

Beyond the first pruning, $\rho$ falls on every scale, reaching 1.92 on large-v2 by L-8, while the absolute gap widens at every depth, from 13.5\,pp unpruned to 15.0, 18.0, 20.6, and 24.5\,pp at L-8 (Table~\ref{tab:wer-eth}). This is the divergence anticipated in Section~\ref{sec:eval-details}: deep pruning drives every group toward high error rates, and the quotient of two large numbers is smaller than that of two small ones even when the difference between them has grown. Our primary measure resolves it, since at L-8 Black speakers face 51.0\% WER against 26.5\% for Asian speakers. Every scale inherits the disparity, then, but only the largest conceals what pruning does to it.

\begin{table}[ht]
\centering
\footnotesize
\setlength{\tabcolsep}{5pt}
\begin{tabular}{lccccc}
\toprule
Model & L-0 & L-2 & L-4 & L-6 & L-8 \\
\midrule
small (12L)    & 2.03 & 1.70 & \textit{1.51} & \textit{1.49} & \textit{1.13} \\
medium (24L)   & 2.16 & 2.01 & 1.94 & 1.97 & \textit{1.66} \\
large-v2 (32L) & 1.99 & \textbf{2.15} & 2.10 & 1.93 & 1.92 \\
\bottomrule
\end{tabular}
\caption{Fair-Speech, disparity ratio $\rho$ between Black and Asian speakers, per encoder scale. Italic values lie beyond the usable range (aggregate WER $>40\%$). Only large-v2 shows a rise at the first prune, where its aggregate WER also improves.}
\label{tab:rho-scale}
\end{table}

\subsection{The effect of LoRA adaptation (RQ3)}
\label{sec:results-lora}

Fig.~\ref{fig:heatmap-pair} summarizes the LoRA comparison: each cell is a subgroup's change in WER relative to the unpruned model (warm colors mean degradation), with the base model in panel (a), the LoRA adapted model in panel (b), and the aggregate in the bottom row; the underlying per-group WERs are in Appendix Table~\ref{tab:app-lora}. The LoRA-adapted model is better everywhere: with the unpruned encoder, aggregate WER drops from 21.6\% to 17.7\%, and at eight removed layers from 37.6\% to 33.0\%. Adaptation also extends the usable pruning range through ten removed layers, two more than the base model.

The disparity, however, does not shrink with the error rate. At every depth, the Black-to-Asian ratio is wider with LoRA than without LoRA: at L-0 it rises from 1.99 (base) to 2.20 (LoRA), at L-2 from 2.15 to 2.36, and at L-8 from 1.93 to 2.22. The reason is visible in panel (b): LoRA adaptation compensates the best-performing groups better. At ten removed layers, it recovers 8.0\,pp of Asian speakers' degradation but only 3.3\,pp of Black speakers'.

\subsection{Beyond Fair-Speech: Common Voice English, Dutch, and Danish}
\label{sec:beyond}

\begin{table}[ht]
\centering
\footnotesize
\setlength{\tabcolsep}{4pt}
\begin{tabular}{lrccccc}
\toprule
Group & $n$ & L-0 & L-2 & L-4 & L-6 & L-8 \\
\midrule
US           & 1,145 & $11.1$ & $12.0$ & $13.9$ & $16.7$ & $17.8$ \\
England      &   358 & $12.2$ & $13.2$ & $15.1$ & $17.1$ & $18.9$ \\
India/S-Asia &   508 & $15.8$ & $17.8$ & $19.1$ & $21.0$ & $24.8$ \\
\midrule
Aggregate    & 3,008 & $12.1$ & $13.0$ & $14.7$ & $17.0$ & $19.1$ \\
\bottomrule
\end{tabular}
\caption{Common Voice 22 English, accent (Whisper large); $n$ is the number of utterances per group. India/South Asia has the highest WER at every depth. Unlike Fair-Speech, the aggregate degrades from the first prune.}
\label{tab:app-cven}
\end{table}

\begin{table}[ht]
\centering
\footnotesize
\setlength{\tabcolsep}{4pt}
\begin{tabular}{lrccccc}
\toprule
Group & $n$ & L-0 & L-2 & L-4 & L-6 & L-8 \\
\midrule
Netherlands & 3,817 & $11.5$ & $12.9$ & $15.0$ & $18.1$ & $20.4$ \\
Belgian     & 1,146 & $13.7$ & $15.2$ & $17.8$ & $20.5$ & $25.2$ \\
\midrule
Aggregate   & 5,306 & $11.8$ & $13.3$ & $15.5$ & $18.5$ & $21.4$ \\
\bottomrule
\end{tabular}
\caption{Common Voice Dutch, accent (Whisper large); $n$ is the number of utterances per group. Belgian speakers have higher WER than Netherlands speakers at every depth.}
\label{tab:app-nl}
\end{table}

\begin{table}[ht]
\centering
\footnotesize
\setlength{\tabcolsep}{4pt}
\begin{tabular}{lrccccc}
\toprule
Group & $n$ & L-0 & L-2 & L-4 & L-6 & L-8 \\
\midrule
Male   & 3,372 & $13.5$ & $15.3$ & $17.7$ & $18.5$ & $20.6$ \\
Female & 1,163 & $10.4$ & $11.1$ & $13.3$ & $14.8$ &  $15.9$ \\
\midrule
Aggregate & 4,535 & $12.7$ & $14.2$ & $15.5$ & $16.5$ & $19.4$ \\
\bottomrule
\end{tabular}
\caption{Common Voice Dutch, gender (Whisper large); $n$ is the number of utterances per group. Male speakers have higher WER at every depth.}
\label{tab:app-nl-gender}
\end{table}

\begin{table}[ht]
\centering

\footnotesize
\setlength{\tabcolsep}{3pt}
\begin{tabular}{llrccccc}
\toprule
Axis & Group & $n$ & L-0 & L-2 & L-4 & L-6 & L-8 \\
\midrule
\multirow{2}{*}{Gender}
 & Male              & 1,031 & $36.0$ & $38.3$ & $40.6$ & $49.0$ & $51.8$ \\
 & Female$^\dagger$  &   299 & $33.0$ & $34.1$ & $35.2$ & $42.9$ & $48.0$ \\
 \midrule
 & Aggregate         & 1,330 & $35.3$ & $37.4$ & $39.4$ & $47.6$ & $50.9$ \\
\midrule
\multirow{6}{*}{Age}
 & Teens$^\dagger$   &    50 & $40.3$ & $42.0$ & $43.6$ & $49.9$ & $53.6$ \\
 & Twenties          &   601 & $41.1$ & $43.2$ & $45.2$ & $53.7$ & $56.3$ \\
 & Thirties          &   392 & $31.5$ & $34.2$ & $36.8$ & $45.0$ & $48.3$ \\
 & Forties$^\dagger$ &   150 & $26.3$ & $27.2$ & $28.0$ & $34.1$ & $36.7$ \\
 & Fifties$^\dagger$ &    98 & $29.2$ & $31.8$ & $34.4$ & $44.5$ & $53.2$ \\
 & Sixties$^\dagger$ &    76 & $36.2$ & $36.4$ & $36.6$ & $44.8$ & $48.7$ \\
 \midrule
 & Aggregate         & 1,367 & $36.5$ & $38.7$ & $40.9$ & $49.2$ & $52.5$ \\
\bottomrule
\end{tabular}
\caption{Common Voice Danish (Whisper large); $n$ is the number of utterances per group. Groups marked $\dagger$ fall below the analysability threshold and are reported for coverage only.}
\label{tab:app-da}
\end{table}

Everything discussed so far comes from Fair-Speech. Fair-Speech provides a comprehensive suite for evaluating multiple demographic disparities. However, it is limited only to the English language. To move beyond English, we also used the demographic labels provided in the CV-22 in different languages. Specifically, we used English, Danish, and Dutch data as in \citet{Kolluri2026OnTR}. This also allows us to determine whether the observed effects generalise across datasets and languages. The results reveal a more nuanced pattern: demographic disparities generally persist under pruning, but the concealed amplification observed on Fair-Speech does not consistently transfer to CV-22.

CV-22 English provides the clearest contrast (Table~\ref{tab:app-cven}). Speakers with Indian and South Asian accents have the highest WER at every pruning depth, and their absolute disadvantage relative to US speakers grows as more layers are removed. However, the relative disparity remains stable: the WER ratio changes only from 1.42 at L-0 to 1.39 at L-8. Moreover, aggregate WER deteriorates from the first pruning step. Thus, pruning reduces overall performance, but it neither disproportionately amplifies the accent gap nor conceals the degradation from a practitioner monitoring aggregate WER.

CV-22 Dutch exhibits the same persistence without amplification (Tables~\ref{tab:app-nl} and~\ref{tab:app-nl-gender}; Fig.~\ref{fig:app-nl}). Belgian speakers consistently have higher WER than speakers from the Netherlands, but the ratio remains between 1.1 and 1.3 across pruning depths. The gender axis follows a similar pattern: male speakers have higher WER at every depth, while the ratio remains close to 1.3. These disparities therefore survive pruning but do not systematically widen.

CV-22 Danish marks the limit of the analysis (Table~\ref{tab:app-da}). Its unpruned aggregate WER is already 35.5\%, and the 40\% usability threshold is crossed after removing four layers. Demographic annotations are also sparse, with most groups—including all female speakers—falling below the analysability threshold. We therefore treat the Danish results as coverage information rather than evidence for group-level effects.

Together, these findings answer RQ2. Pruning did not reduce an existing disparity in any evaluated corpus. However, persistent disparities did not necessarily become amplified: on CV-22 English and Dutch, \emph{relative} gaps remained broadly stable, while aggregate degradation was immediately visible. Concealed disparity amplification occurred only on Fair-Speech with the largest encoder. Evaluating on Fair-Speech does introduce a distribution shift from the Common Voice training data, which may penalize underrepresented linguistic features; however, \emph{absolute} disparities widened under compression in both our in-domain and out-of-domain settings, so this widening cannot be attributed to corpus shift alone. Consequently, the presence of a baseline performance gap cannot predict how compression will affect a group; the effect must be evaluated separately for each corpus, demographic axis, and model scale.

\section{Conclusion}
\label{sec:conclusion}

Demographic bias~\citep{Veliche2024TowardsMF} and layer pruning~\citep{Kolluri2026OnTR} have been studied extensively, but their interaction remains underexplored. In this work, we investigate how pruning affects demographic disparities in ASR. On Fair-Speech, removing two layers from Whisper Large slightly improves aggregate WER, while WER for Black speakers increases. The Black-to-Asian WER ratio rises from 1.99 to 2.15, and after removing eight layers, the absolute gap widens from 13.5 to 24.5 pp. Pruning also increases the low-SES vs affluent gap from 5.9 to 7.2 points and more than triples the gender gap. Thus, aggregate improvements can conceal disproportionate degradation among already disadvantaged groups. LoRA improves aggregate WER at every depth but similarly widens the Black-to-Asian ratio by benefiting better-performing groups more. On CV-22 English, Danish, and Dutch, demographic gaps persist but do not clearly amplify, while sparse annotations and high WER prevent reliable group-level conclusions for Danish.

These findings motivate fairness-aware compression objectives that prioritise worst-group rather than average performance. Future work should examine compression methods that not only focus on improving overall performance but also give equal importance to all demographics.

\section*{Limitations}

This study investigates a single configuration: Whisper is the only speech encoder with three variants, Qwen2.5-3B the only LLM backbone, and top-down layer pruning the only compression technique. All configurations are trained with a single random seed (42). Since the projector is retrained at every pruning depth, seed replication would multiply across the full grid of encoder scales, depths, and languages, which was beyond our compute budget. To mitigate this, we report paired bootstrap significance over the evaluation set and ground our claims in trends that hold consistently across the sweep rather than in differences within any single configuration. Multi-seed replication remains an important direction for future work. The dataset scope is similarly narrow. The main analysis rests on Fair-Speech, an English read-speech corpus with self-reported demographic labels, and the cross-lingual evidence is thin: pruned Danish models were too degraded for group comparison, and the Dutch corpus records only accent, gender, and age. The demographic axes share speakers and are analyzed separately, leaving intersectional effects unexamined. We evaluate no mitigation techniques.

\section*{Acknowledgments}

This work was supported by a Knowledge Transfer Partnership (KTP) project (project number 10131983) funded by UKRI through Innovate UK, in collaboration with Hivedome. RS was supported by the ELOQUENCE project (grant number 101135916) funded by the UKRI and the European Union. Views and opinions expressed are, however, those of the author(s) only and do not necessarily reflect those of the UKRI, European Union, or European Commission-EU. Neither the European Union nor the granting authority can be held responsible for them.

\bibliography{latex/ref}

\newpage

\appendix

\section{Appendix: Training Details}
\label{app:repro}

Table~\ref{tab:epochs} reports the batch size and epoch count for each language. All
other optimisation settings are identical across runs and are given in
Section~\ref{sec:impl}.

Epoch counts scale inversely with corpus size: English (100 hours) converges within two passes, while Dutch (54 hours) and Danish (4.2 hours) require more passes to converge on their smaller corpora. LoRA configurations are trained for roughly twice as many epochs as projector-only runs, since the adapters are optimised jointly with a projector that is itself trained from scratch. Batch size is reduced to 4 for Danish, whose corpus is too small to fill larger batches without excessive repetition within an epoch. All settings are held constant across encoder scales and pruning depths, so configurations differ only in encoder capacity.

\begin{table}[ht]
\centering\small
\begin{tabular}{lccc}
\toprule
Language & Batch & Proj. & LoRA \\
\midrule
English & 8 & 2 & 4 \\
Dutch & 8 & 4 & 8 \\
Danish & 4 & 6 & 8 \\
\bottomrule
\end{tabular}
\caption{Batch size and epoch counts per language. Settings are identical across all
three encoder scales and all pruning depths.}
\label{tab:epochs}
\end{table}

\section{Appendix: Relative-to-baseline WER on Fair-Speech}
\label{app:rel}

Tables~\ref{tab:rel-eth} and~\ref{tab:rel-scale} report WER at each pruning depth relative to the unpruned model (WER$_{Lx}$/WER$_{L0}$): per ethnic group on Whisper large, and across model scales for the aggregate and the two headline groups. Ratios below 1.00 mean improvement over the unpruned model.

\begin{table}[ht]
\centering
\footnotesize
\setlength{\tabcolsep}{5pt}
\begin{tabular}{lrccccc}
\toprule
Group & $n$ & L-0 & L-2 & L-4 & L-6 & L-8 \\
\midrule
Asian       & 3,854 & 1.00 & 0.95 & 1.19 & 1.62 & 1.93 \\
Native Haw. &   969 & 1.00 & 1.02 & 1.33 & 1.84 & 2.34 \\
Hispanic    & 2,811 & 1.00 & 0.96 & 1.12 & 1.37 & 1.69 \\
White       & 5,619 & 1.00 & 0.98 & 1.10 & 1.38 & 1.54 \\
Native Am.  & 4,616 & 1.00 & 0.88 & 1.03 & 1.35 & 1.56 \\
MENA        &   749 & 1.00 & 0.97 & 1.13 & 1.49 & 1.63 \\
Black       & 7,799 & 1.00 & \textbf{1.03} & 1.26 & 1.57 & 1.87 \\
\midrule
Aggregate   & 26,417 & 1.00 & \textbf{0.98} & 1.17 & 1.48 & 1.74 \\
\bottomrule
\end{tabular}
\caption{Fair-Speech, WER relative to each group's own unpruned baseline (Whisper large); $n$ is the number of utterances per group. Values below 1.00 indicate improvement. At L-2 the aggregate is at 0.98 while Black speakers are at 1.03, the highest of any group.}
\label{tab:rel-eth}
\end{table}

\begin{table}[ht]
\centering
\footnotesize
\setlength{\tabcolsep}{1.5pt}
\begin{tabular}{llrccccc}
\toprule
Model & Series & $n$ & L-0 & L-2 & L-4 & L-6 & L-8 \\
\midrule
\multirow{3}{*}{large-v2 (32L)}
  & Asian     &  3,854 & 1.00 & 0.95 & 1.19 & 1.62 & 1.93 \\
  & Black     &  7,799 & 1.00 & \textbf{1.03} & 1.26 & 1.57 & 1.87 \\
  & Aggregate & 26,417 & 1.00 & \textbf{0.98} & 1.17 & 1.48 & 1.74 \\
\midrule
\multirow{3}{*}{medium (24L)}
  & Asian     &  3,854 & 1.00 & 1.17 & 1.51 & 1.72 & \textit{2.64} \\
  & Black     &  7,799 & 1.00 & 1.09 & 1.35 & 1.57 & \textit{2.03} \\
  & Aggregate & 26,417 & 1.00 & 1.11 & 1.36 & 1.52 & \textit{2.10} \\
\midrule
\multirow{3}{*}{small (12L)}
  & Asian     &  3,854 & 1.00 & 1.62 & \textit{2.56} & \textit{2.90} & \textit{5.14} \\
  & Black     &  7,799 & 1.00 & 1.35 & \textit{1.91} & \textit{2.12} & \textit{2.85} \\
  & Aggregate & 26,417 & 1.00 & 1.41 & \textit{2.12} & \textit{2.37} & \textit{3.69} \\
\bottomrule
\end{tabular}
\caption{Fair-Speech, WER relative to each model's own unpruned baseline; $n$ is the number of utterances. Italic values lie beyond the usable range (aggregate WER $>40\%$). Only large-v2 has a depth where the aggregate improves while Black speakers degrade.}
\label{tab:rel-scale}
\end{table}

\section{Appendix: Per-group WER at small and medium scale}
\label{app:scale-tables}

\begin{table}[t]
\centering
\footnotesize
\setlength{\tabcolsep}{4pt}
\begin{tabular}{lrccccc}
\toprule
Group & $n$ & L-0 & L-2 & L-4 & L-6 & L-8 \\
\midrule
Asian       & 3,854 & $12.5$ & $14.7$ & $18.8$ & $21.4$ & \textit{32.9} \\
Native Haw. &   969 & $14.8$ & $17.6$ & $20.3$ & $24.1$ & \textit{36.1} \\
White       & 5,619 & $18.5$ & $21.7$ & $24.9$ & $28.2$ & \textit{38.9} \\
Hispanic    & 2,811 & $18.7$ & $21.9$ & $25.0$ & $26.0$ & \textit{39.4} \\
Native Am.  & 4,616 & $19.3$ & $22.7$ & $26.0$ & $26.5$ & \textit{38.5} \\
MENA        &   749 & $22.8$ & $25.5$ & $28.2$ & $33.1$ & \textit{42.7} \\
Black       & 7,799 & $27.0$ & $29.5$ & $36.4$ & $42.2$ & \textit{54.6} \\
\midrule
Aggregate   & 26,417 & $20.4$ & $22.7$ & $27.7$ & $31.0$ & \textit{42.9} \\
\bottomrule
\end{tabular}
\caption{Fair-Speech, ethnicity (Whisper medium); $n$ is the number of
utterances per group. Italic values lie beyond the usable range
(aggregate WER $>40\%$).}
\label{tab:app-eth-medium}
\end{table}

\begin{table}[t]
\centering
\footnotesize
\setlength{\tabcolsep}{4pt}
\begin{tabular}{lrccccc}
\toprule
Group & $n$ & L-0 & L-2 & L-4 & L-6 & L-8 \\
\midrule
Asian       & 3,854 & $17.9$ & $28.9$ & \textit{45.8} & \textit{51.9} & \textit{91.9} \\
Native Haw. &   969 & $18.7$ & $31.2$ & \textit{51.8} & \textit{60.9} & \textit{101.8} \\
White       & 5,619 & $23.3$ & $32.9$ & \textit{51.4} & \textit{56.9} & \textit{94.0} \\
Native Am.  & 4,616 & $23.6$ & $34.0$ & \textit{53.5} & \textit{60.5} & \textit{100.8} \\
Hispanic    & 2,811 & $24.5$ & $33.0$ & \textit{51.6} & \textit{57.8} & \textit{98.6} \\
MENA        &   749 & $30.8$ & $39.2$ & \textit{55.9} & \textit{61.8} & \textit{103.4} \\
Black       & 7,799 & $36.3$ & $49.1$ & \textit{69.3} & \textit{77.1} & \textit{103.6} \\
\midrule
Aggregate   & 26,417 & $26.8$ & $37.6$ & \textit{56.6} & \textit{63.4} & \textit{98.8} \\
\bottomrule
\end{tabular}
\caption{Fair-Speech, ethnicity (Whisper small); $n$ is the number of utterances per group. Italic values lie beyond the usable range (aggregate WER $>40\%$).}
\label{tab:app-eth-small}
\end{table}

Tables~\ref{tab:app-eth-small} and~\ref{tab:app-eth-medium} give the per-group WER behind Fig.~\ref{fig:scale}, in the same form as Table~\ref{tab:wer-eth} for large-v2. Black speakers have the highest WER at every scale and depth.

\section{Appendix: Fair-Speech coverage, SES, gender, and age}
\label{app:fs-axes}

\begin{table}[t]
\centering

\footnotesize
\begin{tabular}{llrr}
\toprule
Axis & Group & $n$ & Minutes \\
\midrule
\multirow{7}{*}{Ethnicity}
 & Black       &  7,799 & 1,010 \\
 & White       &  5,619 &   702 \\
 & Native Am.  &  4,616 &   556 \\
 & Asian       &  3,854 &   462 \\
 & Hispanic    &  2,811 &   372 \\
 & Native Haw. &    969 &    94 \\
 & MENA        &    749 &    86 \\
\midrule
\multirow{3}{*}{SES}
 & Low      & 14,741 & 1,837 \\
 & Medium   &  9,809 & 1,219 \\
 & Affluent &  1,867 &   224 \\
\midrule
\multirow{2}{*}{Gender}
 & Female & 14,382 & 1,821 \\
 & Male   & 12,035 & 1,459 \\
\midrule
\multirow{4}{*}{Age}
 & 31--45 & 12,755 & 1,493 \\
 & 46--65 &  5,653 &   841 \\
 & 23--30 &  4,164 &   530 \\
 & 18--22 &  3,845 &   416 \\
\bottomrule
\end{tabular}
\caption{Fair-Speech coverage per demographic group: utterances and
minutes of audio. Every analyzed group exceeds the analysability
threshold of 200 utterances and 30 minutes.}
\label{tab:app-fs-coverage}
\end{table}

\begin{figure}[t]
\centering
\includegraphics[width=\columnwidth]{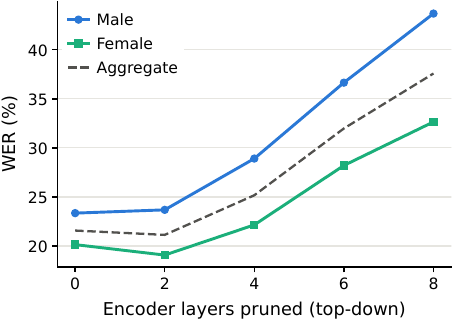}
\caption{WER by gender under top-down pruning (Fair-Speech, Whisper
large-v2; dashed = aggregate). Male speakers trail at every depth, and
the gap widens from 3.3\,pp at L-0 to 11.1\,pp at L-8: female speakers
improve slightly at the first prune while male speakers do not, and they
degrade more slowly thereafter.}
\label{fig:app-gender}
\end{figure}

\begin{figure}[t]
\centering
\includegraphics[width=\columnwidth]{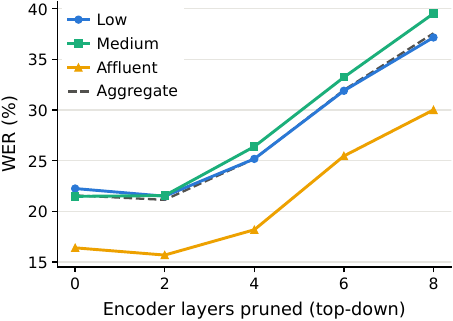}
\caption{WER by socioeconomic group under top-down pruning (Fair-Speech, Whisper large; dashed = aggregate). Affluent speakers keep an advantage at every depth. The gap does not amplify: the Low-to-Affluent ratio falls from 1.36 to 1.24 as all groups degrade. 
}
\label{fig:ses}
\end{figure}

\begin{table}[t]
\centering
\footnotesize
\setlength{\tabcolsep}{3pt}
\begin{tabular}{llrccccc}
\toprule
Axis & Group & $n$ & L-0 & L-2 & L-4 & L-6 & L-8 \\
\midrule
\multirow{3}{*}{SES}
 & Low       & 14,741 & $22.3$ & $21.5$ & $25.2$ & $31.9$ & $37.2$ \\
 & Medium    &  9,809 & $21.5$ & $21.6$ & $26.4$ & $33.2$ & $39.5$ \\
 & Affluent  &  1,867 & $16.4$ & $15.7$ & $18.2$ & $25.5$ & $30.0$ \\
\midrule
\multirow{2}{*}{Gender}
 & Male      & 12,035 & $23.4$ & $23.7$ & $28.9$ & $36.6$ & $43.7$ \\
 & Female    & 14,382 & $20.1$ & $19.1$ & $22.2$ & $28.2$ & $32.6$ \\
\midrule
\multirow{4}{*}{Age}
 & 18--22    &  3,845 & $16.2$ & $15.5$ & $18.8$ & $25.8$ & $31.0$ \\
 & 23--30    &  4,164 & $19.3$ & $18.1$ & $22.0$ & $28.3$ & $34.5$ \\
 & 31--45    & 12,755 & $21.2$ & $21.3$ & $26.5$ & $34.3$ & $41.0$ \\
 & 46--65    &  5,653 & $27.0$ & $26.1$ & $28.5$ & $33.7$ & $36.9$ \\
\midrule
 & Aggregate & 26,417 & $21.6$ & $21.1$ & $25.2$ & $32.0$ & $37.6$ \\
\bottomrule
\end{tabular}
\caption{Fair-Speech, socioeconomic status, gender, and age (Whisper large); $n$ is the number of utterances per group.}
\label{tab:app-axes}
\end{table}

\begin{figure}[t]
\centering
\includegraphics[width=\columnwidth]{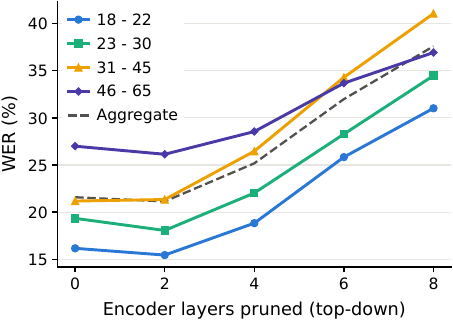}
\caption{WER by age group under top-down pruning (Fair-Speech, Whisper
large-v2; dashed = aggregate). The oldest group (46--65), worst at
baseline, degrades slowest ($+9.9$\,pp by L-8), while the 31--45 group
degrades fastest ($+19.8$\,pp) and overtakes it as the worst-performing
group by L-6. The worst-best gap narrows, but by leveling down, not by
helping the group that started worst.}
\label{fig:app-age}
\end{figure}

Table~\ref{tab:app-fs-coverage} reports the number of utterances and
minutes of audio behind every Fair-Speech group;
Table~\ref{tab:app-axes} reports WER for the remaining demographic axes,
summarized in Fig.~\ref{fig:gaps}; Figs.~\ref{fig:app-gender}
and~\ref{fig:app-age} show the gender and age axes in detail.

\section{Appendix: Per-group WER under LoRA adaptation}
\label{app:lora}

\begin{table}[t]
\centering
\footnotesize
\setlength{\tabcolsep}{3pt}
\begin{tabular}{lrcccccc}
\toprule
Group & $n$ & L-0 & L-2 & L-4 & L-6 & L-8 & L-10 \\
\midrule
Asian       & 3,854 & $11.1$ & $11.2$ & $14.2$ & $18.5$ & $22.2$ & $29.1$ \\
Native Haw. &   969 & $13.8$ & $14.4$ & $16.0$ & $21.3$ & $27.0$ & $36.7$ \\
Native Am.  & 4,616 & $14.4$ & $15.6$ & $18.0$ & $22.2$ & $26.2$ & $34.8$ \\
Hispanic    & 2,811 & $15.6$ & $16.1$ & $19.6$ & $22.8$ & $28.6$ & $33.5$ \\
White       & 5,619 & $16.6$ & $17.9$ & $18.8$ & $23.8$ & $25.8$ & $32.9$ \\
MENA        &   749 & $17.9$ & $21.0$ & $23.1$ & $27.6$ & $30.6$ & $40.0$ \\
Black       & 7,799 & $24.4$ & $26.4$ & $32.5$ & $41.7$ & $49.3$ & $55.3$ \\
\midrule
Aggregate   & 26,417 & $17.7$ & $18.9$ & $22.3$ & $28.2$ & $33.0$ & $40.0$ \\
\bottomrule
\end{tabular}
\caption{Fair-Speech, ethnicity under LoRA adaptation (Whisper large);
$n$ is the number of utterances per group. WER is lower than in the base
model (Table~\ref{tab:wer-eth}) for every group at every depth.}
\label{tab:app-lora}
\end{table}

Table~\ref{tab:app-lora} reports per-group WER under LoRA adaptation
(Fair-Speech, Whisper large), complementing
Fig.~\ref{fig:heatmap-pair}; compare with the base model in
Table~\ref{tab:wer-eth}. Per-utterance outputs were not retained for the
LoRA runs; SD here is estimated from stored resampling intervals and
reflects the precision of the group WER rather than per-utterance spread.

\section{Appendix: Common Voice English, Dutch, and Danish}
\label{app:othercorpora}

\begin{figure}[!htb]
\centering
\includegraphics[width=\columnwidth]{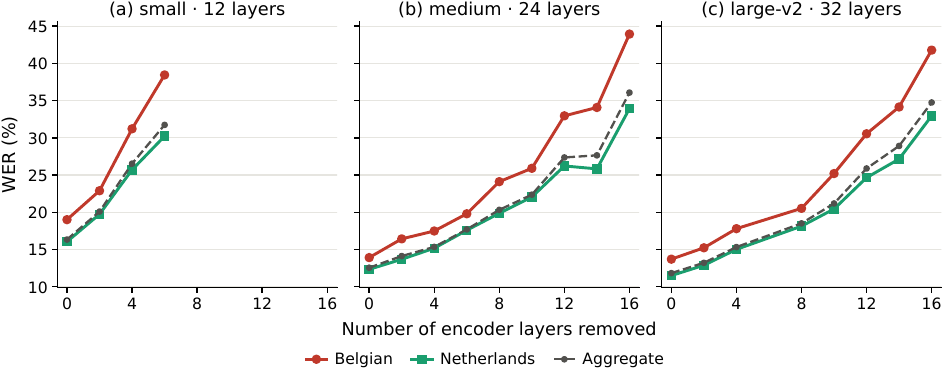}
\caption{Common Voice Dutch: Belgian vs Netherlands accent WER under
top-down pruning, per Whisper scale (usable range only; dashed =
aggregate). Belgian speakers trail at every scale and depth, but the ratio
stays between 1.1 and 1.3 throughout: the gap does not amplify. Each panel
ends where its model leaves the usable range.}
\label{fig:app-nl}
\end{figure}

This section supports the cross-corpus summary in Section~\ref{sec:beyond}.
Accent groups shown are those meeting the analysability threshold
($\ge$200 utterances and $\ge$30 minutes).

Common Voice English (Table~\ref{tab:app-cven}) offers a useful contrast
to Fair-Speech. The accent hierarchy is stable, with India/South Asia
trailing US and England at every depth, and the absolute gap grows as
pruning deepens. But there is no concealment window: the aggregate
degrades from the very first prune, so a practitioner watching only the
corpus WER would already see the cost. The relative gap does not widen
either: 1.42 at L-0, 1.39 at L-8.

Dutch shows the same persistence without amplification
(Tables~\ref{tab:app-nl} and~\ref{tab:app-nl-gender},
Fig.~\ref{fig:app-nl}). Belgian speakers trail Netherlands speakers at
every scale and every usable depth, yet the ratio between them stays
essentially flat, between 1.1 and 1.3, across small, medium, and
large-v2 alike. The Dutch gender axis behaves the same way: male
speakers trail at every depth, echoing the Fair-Speech gender direction,
with the ratio flat near 1.3. The inherited gaps survive pruning
unchanged; unlike the racial gap on Fair-Speech, they do not widen, and
they behave the same at every model scale.

Danish marks the edge of what this analysis can reach
(Table~\ref{tab:app-da}). Even unpruned, the aggregate WER sits at
35.5\%, leaving little headroom: the 40\% bound is crossed by six removed
layers. The demographic annotations are also sparse; most groups fall
below the analysability threshold, including all female speakers. We
therefore draw no group-level conclusions for Danish, and the table
documents coverage rather than findings. Together, the three corpora
bound the Fair-Speech result: disparities persist everywhere, but
amplification hidden under an improving aggregate appeared only on
Fair-Speech with the largest encoder.

\end{document}